\documentclass[a4paper, 11pt]{article}

\usepackage{graphicx}
\usepackage[all]{xy}
\usepackage{amsfonts}
\usepackage{dsfont}
\usepackage{mathrsfs}
\usepackage{amssymb}
\usepackage{bbm}
\usepackage{epsfig}
\usepackage{amsmath}
\usepackage{appendix}
\usepackage{color}
\usepackage[english]{babel}
\usepackage[pdftex, a4paper]{geometry}
\usepackage{fancyhdr}
\def\Reals{\mathop{\hbox{\mit I\kern-.2em R}}\nolimits}

\def\Complexes{{\hbox{\mit C\kern-.46em
            \vrule depth 0ex height 1.4ex width .05em\kern.41em}}}

\usepackage[english]{babel}
\usepackage{epsfig}
\usepackage{dsfont}
\usepackage{mathrsfs}
\usepackage[mathscr]{eucal}
\usepackage{graphicx}
\usepackage{amsmath,amsfonts,amsthm,amssymb,amsbsy,amsopn,amstext}
\usepackage{latexsym}
\usepackage{color,multicol,makeidx}
\usepackage{CJK}
\usepackage{indentfirst}
\usepackage{bm}
\usepackage{mathrsfs}
\usepackage{float}
\usepackage[caption=false,font=normalsize,labelfont=sf,textfont=sf]{subfig}
\usepackage{multirow}

\newtheorem{thm}{Theorem}
\newtheorem{defn}{Definition}

\newtheorem{lem}{Lemma}
\newtheorem{remark}{Remark}
\newtheorem{prop}{Proposition}

\title{\bf Emergent Behaviors  over Signed Random Dynamical Networks: State-Flipping  Model}
\date{}
\author{Guodong Shi, Alexandre Proutiere, Mikael  Johansson, \\
John S. Baras, and Karl H. Johansson}
\begin{document}
\maketitle

\begin{abstract}
Recent studies from social, biological, and engineering network systems have drawn attention to the dynamics over signed networks, where each link is associated with a positive/negative sign indicating trustful/mistrustful, activator/inhibitor,  or secure/malicious interactions.   We study asymptotic dynamical patterns that emerge among a set of nodes that interact in a
dynamically evolving signed random network.  Node  interactions take place at random on a sequence of deterministic signed graphs.  Each node receives positive or negative  recommendations from its neighbors depending on the sign of the interaction arcs, and updates its state accordingly. Recommendations along a positive arc follow the standard consensus update. As in the work by  Altafini, negative  recommendations use an update where the sign of  the neighbor state is flipped. Nodes may weight positive and negative recommendations differently, and random processes are introduced to model the time-varying attention that nodes pay to these recommendations. Conditions for almost sure convergence and  divergence  of the node states are established. We show that under this so-called state-flipping model, all links contribute to a consensus of the absolute values of the nodes, even under switching sign patterns and dynamically changing environment. A no-survivor property is established, indicating that every node state diverges almost surely if the maximum network state diverges.
\end{abstract}


\section{Introduction}
\subsection{Motivation}
The need to model, analyze and engineer large complex networks appears  in a wide spectrum of scientific disciplines, ranging from social sciences and biology to physics and engineering \cite{degroot,vic95,jad03}. In many cases, these networks are composed of relatively simple agents that interact locally with their neighbors based on a very limited knowledge about the system state.
Despite the simple local interactions, the resulting networks can display a rich set of emergent behaviors, including certain forms of intelligence and learning~\cite{swarm,naive}.


Consensus problems, in which the aim is to compute a weighted average of the initial values held by a collection of nodes, play a fundamental role in the study of node dynamics over complex networks.   Early  work \cite{degroot} focused on understanding how opinions evolve in a network of agents, and showed that a simple deterministic opinion update based on the mutual trust and the differences in belief between interacting agents could lead to global convergence of the beliefs. Consensus dynamics has since then been widely adopted for describing opinion dynamics in social networks, e.g., \cite{naive,social2,misinformation}. In engineering sciences, a huge amount of literature has studied these algorithms for distributed averaging, formation  forming and load balancing between collaborative agents under fixed or time-varying interaction networks \cite{tsi,xiao,juliencdc,mor,ren,saber,caoming,julien13}. Randomized  consensus seeking has also been widely studied, motivated by  the random nature of interactions and updates in real complex networks~\cite{hatano,boyd,fagnani,kar2,touri,jad08,it10,mateisiam,TON13}.

Interactions in large-scale networks are not always collaborative since nodes  take on different, or even opposing, roles. A convenient framework for modeling different roles and relationships between agents is to use signed graphs. Signed graphs were introduced in the classical work by Heider in 1946~\cite{heider46} to model the structure of social networks, where a positive link represents a friendly relation between two agents, and a negative link an unfriendly one.  In \cite{galam96}, a dynamic model based on a signed graph with positive links between nodes (representing nations) belonging to the same coalition and negative otherwise,  was introduced to study  the stability of  world politics. In biology, sign patterns have been used to describe activator--inhibitor interactions between pairs of chemicals \cite{biologymath},  neural networks for vision and learning \cite{adptive}, and gene regulatory networks~\cite{nature13}. In all these examples, the state updates that happen when two nodes interact depend on the sign of the arc between the nodes in the underlying graph. The understanding of the emergent dynamical behaviors in  networks with agents having different roles is much more limited than our knowledge about collaborative agents performing consensus algorithms.


It is intriguing to investigate  what happens when two types of dynamics are coupled in a single network. Naturally we ask: how should we model the dynamics of positive and negative interactions? When do behaviors such as consensus, swarming and clustering emerge, and how does the structure of the sign patterns influence these behaviors? In this paper, continuing the previous efforts in \cite{altafini13,shiJSAC}, we answer these questions for a general model of opinion formation in dynamic signed random networks.

\subsection{Contributions}

In this paper, we study  a scheme of randomized node interaction over a signed network of nodes, and show how the nodes' states asymptotically evolve under these positive or negative interactions.   A sequence of deterministic signed graphs defines the dynamics of the network. Random node interactions take place under independent,  but not necessarily identically distributed,  random sampling of the environment.  Once interaction relations have been realized, each node receives a positive recommendation consistent with the standard consensus algorithm from its positive neighbors. Nodes receive negative recommendations from its negative neighbors. {  In this paper we investigate a model where neighbors construct negative recommendations by flipping the sign of their true state during the interaction. This definition of negative interaction was introduced in \cite{altafini13}. After receiving these recommendations, each node puts a (deterministic) weight to each recommendation, and then encodes these weighted recommendations in its state update through stochastic attentions defined by two Bernoulli random variables.

Our model is general, and covers many of the existing node interaction models, e.g., consensus over Erd\H{o}s-R\'{e}nyi graph \cite{hatano}, pairwise randomized gossiping \cite{boyd},  random link failure \cite{kar2}, etc. We allow the sign of each link to be time-varying as well in a dynamically changing environment. We establish conditions for  almost sure convergence and divergence of the node states. We show that under the state-flipping model, all links contribute to a consensus of the absolute values of the nodes, even under  switching sign patterns. We also show that strong structural balance~\cite{harary56} is crucial for belief clustering, which is consistent with the results derived in \cite{altafini13}.  In the almost sure divergence analysis, we establish that the deterministic weights  nodes put on negative recommendations   play a crucial role in driving the divergence of the network. A no-survivor property is established indicating that every node state diverges almost surely given that the maximum network state diverges. Our analysis does not rely on a spectrum analysis as that used in \cite{altafini13}, but instead  we study the asymptotic behavior of the node states  using a sample-path analysis.

 }


\subsection{Organization}
In Section 2, we present the network dynamics and the node update rules. The state-flipping model is defined  for the negative recommendations. Section 3 presents our main results on the state-flipping model and  the detailed proofs are presented in Section~4. Finally some concluding remarks are drawn in Section 5.
\vspace{3mm}
\subsection*{Notation}

A simple directed graph (digraph) $\mathcal
{G}=(\mathcal {V}, \mathcal {E})$ consists of a finite set
$\mathcal{V}$ of nodes and an arc set
$\mathcal {E}\subseteq \mathcal{V}\times\mathcal{V}$, where  $e=(i,j)\in\mathcal {E}$ denotes   an
{\it arc}  from node $i\in \mathcal{V}$  to $j\in\mathcal{V}$ with $(i,i)\notin \mathcal{E}$ for all $i\in\mathcal{V}$.  We say that node $j$ is {\it reachable} from node $i$ if there is a directed path from $i$ to $j$, with the additional convention that every node is reachable from itself. A node $v$ from which every  node in $\mathcal{V}$ is reachable is called a {\it center node} (or a root). A digraph $\mathcal{G}$ is {\it strongly connected} if every two  nodes are mutually reachable;  $\mathcal{G}$ has a spanning tree if it has a center node; $\mathcal{G}$ is {\it weakly connected} if  a connected undirected graph can be obtained by removing all the directions of the arcs in $\mathcal{E}$. A subgraph   of $\mathcal
{G}=(\mathcal {V}, \mathcal {E})$, is a graph on the same node set $\mathcal {V}$ whose arc set is a subset of $\mathcal {E}$. The induced graph of $\mathcal{V}_i \subseteq \mathcal{V}$ on $\mathcal{G}$, denoted $\mathcal{G}|_{\mathcal{V}_i}$, is the graph $(\mathcal{V}_i, \mathcal{E}_i)$ with $\mathcal{E}_i=(\mathcal{V}_i\times \mathcal{V}_i)\cap \mathcal{E}$. A weakly connected component of $\mathcal
{G}$ is a maximal weakly connected induced graph of $\mathcal
{G}$. If each arc $(i,j)\in\mathcal{E}$ is associated with a sign, either '$+$' or '$-$', $\mathcal {G}$ is called a signed graph and the sign of $(i,j)\in\mathcal {E}$ is denoted  as $\sigma_{ij}$.  The positive and negative subgraphs containing the positive and negative arcs of $\mathcal{G}$, are denoted as $\mathcal {G}^{+}=(\mathcal{V}, \mathcal{E}^+)$ and $\mathcal {G}^-=(\mathcal{V}, \mathcal{E}^-)$, respectively.

Depending on the argument, $|\cdot|$ stands for the absolute value of a real number, the Euclidean norm of a vector or the cardinality of a set. The $\sigma$-algebra generated by a random variable $X$ is denoted as $\sigma(X)$.


\section{Random Network Model  and Node Updates}

We consider a dynamic  network where each node holds and updates its  belief or state when interacting with other nodes. In this section, we present a general model specifying the network dynamics and the way nodes interact.

\subsection{Dynamic Signed Graphs}

We consider a network  with a set $\mathcal{V}=\{1,\dots,n\}$ of $n$ nodes, with $n\geq3$. Time is slotted, and at each slot $t=0,1,\ldots$, each node can interact with its neighbors in a simple  directed graph $\mathcal {G}_t=(\mathcal{V},\mathcal{E}_t)$. The graph evolves over time in an arbitrary and deterministic manner. We assume ${\cal G}_t$ is a signed graph, and we denote by $\sigma_{ij}(t)$ the sign of arc $(i,j)\in \mathcal{E}_t$. The sign of arc $(i,j)$ indicates whether $i$ is a friend ($\sigma_{ij}(t)=+$), or an enemy ($\sigma_{ij}(t)=-$) of node $j$. The positive and negative subgraphs containing the positive and negative arcs of $\mathcal{G}_t$, are denoted by $\mathcal {G}^{+}_t=(\mathcal{V}, \mathcal{E}^+_t)$ and $\mathcal {G}^-_t=(\mathcal{V}, \mathcal{E}^-_t)$, respectively. We say that the sequence of graphs $\{\mathcal {G}_t\}_{t\ge 0}$ is {\it sign consistent} if the sign of any arc $(i,j)$ does not evolve over time, i.e., if for any $s,t \ge 0$,
$$
(i,j)\in \mathcal {E}_s\cap \mathcal {E}_t\  \Longrightarrow \sigma_{ij}(s)=\sigma_{ij}(t).
$$
We also define $\mathcal {G}_\ast=(\mathcal{V},\mathcal{E}_\ast)$ with $\mathcal{E}_\ast=\bigcup_{t=0}^\infty\mathcal{E}_t$ as the total graph of the network. If  $\{\mathcal {G}_t\}_{t\ge 0}$ is sign consistent, then the sign of  each arc $\mathcal{E}_\ast$ never changes and in that case, $\mathcal {G}_\ast=(\mathcal{V},\mathcal{E}_\ast)$ is a well-defined signed graph.

\begin{remark}
 Note that  $\mathcal{G}_t$ is defined over  directed graphs. The only requirement on $\mathcal {G}^+_t$ and $\mathcal {G}^-_t$ is that they should be disjoint, so the signed graph model under consideration is quite general. In particular, we allow that the two possible edge directions coexist between  pair of nodes and that the two directions can have different signs.
\end{remark}

\begin{figure}[t]
\begin{center}
\includegraphics[height=1.6in]{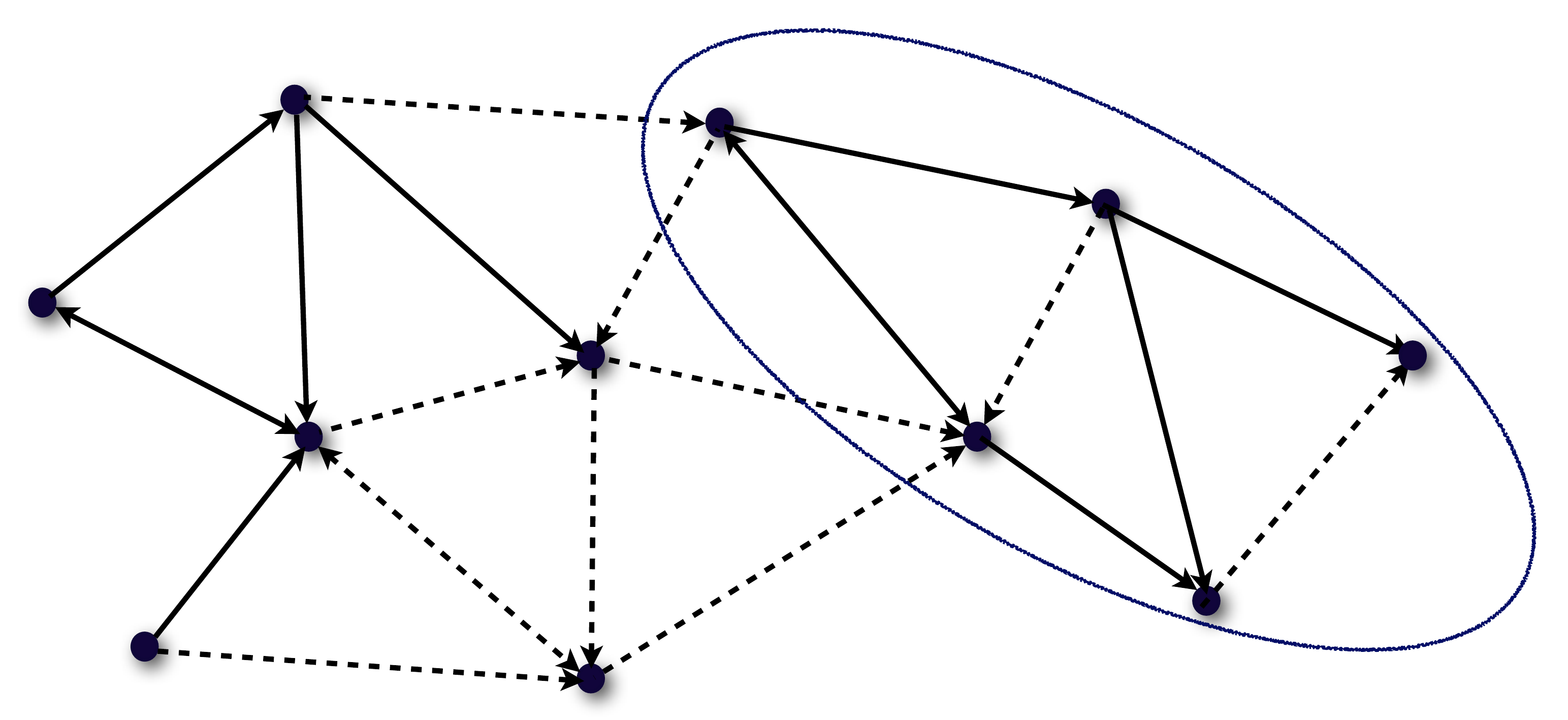}
\caption{A signed network and one of its three positive clusters encircled. The  positive arcs are solid, and the negative arcs are dashed. Note that negative arcs are allowed within each positive cluster. } \label{graph}
\end{center}
\end{figure}

Next we introduce the notion of positive cluster in a signed digraph, which will play an important role in the analysis of the belief dynamics (see Fig. \ref{graph}).

\medskip
\begin{defn} Let $\mathcal {G}$ be a signed digraph with positive subgraph $\mathcal {G}^+$.
 A subset $\mathcal {V}_\ast$ of the set of nodes $\mathcal {V}$ is a  positive cluster  if $\mathcal {V}_\ast$ constitutes a weakly connected component of $\mathcal {G}^{+}$. A positive cluster partition of $\mathcal {G}$ is a partition of $\mathcal{V}$ into $T_p\geq1$ positive clusters $\mathcal{V}_i,i=1,\dots,T_p$, such that $\mathcal{V}=\bigcup_{i=1}^{{\rm T}_{\rm p}} \mathcal{V}_i$.
 \end{defn}
 \medskip

Note that negative arcs may exist between the nodes of a positive cluster. Therefore, a positive-cluster partition of ${\cal G}$ can be seen as an extension of the classical notion of weak structural balance for which negative links are strictly forbidden inside each positive cluster \cite{davis63}.  From the above definition, it is clear that for any signed graph $\mathcal {G}$, there is a unique positive cluster partition $\mathcal{V}=\bigcup_{i=1}^{{\rm T}_{\rm p}} \mathcal{V}_i$, where ${\rm T}_{\rm p}$ is the number of maximal positive clusters covering the entire set  of nodes.

\subsection{Random Interactions}

Time is discrete and at time $t$, node $i$ may only interact with its neighboring nodes in ${\cal G}_t$. We consider a general model for the random node interactions.  At time $t$, some pairs of nodes are randomly selected for interaction. We denote by $E_t\subset {\cal E}_t$ the random subset of arcs corresponding to interacting node pairs at time $t$. More precisely, $E_t$ is sampled from the distribution $\mu_t$ defined over the set $\Omega_t$ of all subsets of arcs in ${\cal E}_t$. We assume that $E_0,E_1,\ldots$ form a sequence of independent sets of arcs. Formally, we introduce the probability space $(\Theta,\mathcal{F},\mathrm{P})$ obtained by taking the product of the probability spaces $(\Omega_t,{\cal S}_t,\mu_t)$, where $\mathcal{S}_t$ is the discrete $\sigma$-algebra on $\Omega_t$: $\Theta=\prod_{t\ge 0}\Omega_t$, ${\cal F}$ is the product of $\sigma$-algebras ${\cal S}_t$, $t\ge 0$, and $\mathrm{P}$ is the product probability measure of $\mu_t, t\ge 0$. We  denote by $G_t=(\mathcal{V}, E_t)$ the random subgraph of ${\cal G}_t$ corresponding to the random set $E_t$ of arcs. The disjoint  sets $E^+_t$ and $E^-_t$ denote the positive and negative arc sets of $E_t$, respectively. Finally, we split the random set of nodes interacting with node $i$ at time $t$ depending on the sign of the corresponding arc: for node $i$, the set of positive neighbors is defined as ${N}^+_i(t):= \big\{j: (j,i)\in E_t^+\big\}$, whereas similarly, the set of negative neighbors is ${N}^-_i(t):= \big\{j: (j,i)\in E_t^-\big\}$.

{
\begin{remark}
The above model is quite general. It includes as special cases  the classical Erd\H{o}s-R\'{e}nyi random graph \cite{er}, gossiping models where a single pair of nodes is chosen at random for interaction \cite{boyd,misinformation}, or where all nodes interact with their neighbors at a given time \cite{fagnani,kar2,degroot,naive}. Independence is the only hard requirement in our random graph process, which is imposed in most existing works on randomized consensus dynamics, e.g.,  \cite{misinformation,fagnani, kar2, boyd}. Non-independent random graph models for randomized consensus were discussed in \cite{it10,mateisiam,shiIT,jad12}.
\end{remark}
}
\subsection{Node updates}

Next we explain how nodes update their states. Each node $i$ holds a state $s_i(t)\in \mathds{R}$ at $t=0,1,\dots$. To update its state at time $t$, node $i$ considers recommendations received from positive and negative neighbors:
\begin{itemize}
 \item[(i)] The positive  recommendation node $i$ receives  at time $t$ is
$$
h_i^+(t):=-\sum_{j\in{N}_i^+(t) } \big(s_i(t)-s_j(t)\big);
$$
\item[(ii)] The negative recommendation node  $i$ receives at time $t$ is  $$
h_i^-(t):=- \sum_{j\in{N}_i^-(t) } \big(s_i(t)+s_j(t)\big).
$$
\end{itemize}
In the above expressions, we use the convention  that summing over empty sets yields a recommendation equal to zero, e.g., when node $i$ has no positive neighbors, then $h_i^+(t)=0$.

Now let $\{{B}_t\}_{t\ge 0}$ and $\{{D}_t\}_{t\ge 0}$ be two sequences of independent Bernoulli random variables. We further assume that $\{{B}_t\}_{t\ge 0}$, $\{{D}_t\}_{t\ge 0}$, and $\{G_t\}_{t\ge 0}$ define independent processes. For any $t\ge 0$, define $b_t=\mathds{E}\{B_t\}$ and $d_t=\mathds{E}\{D_t\}$. The processes $\{{B}_t\}_{t\ge 0}$ and $\{{D}_t\}_{t\ge 0}$ represent the attention that node $i$ pays to the positive and negative recommendations, respectively.

Node $i$ updates its state as
\begin{align}\label{9}
 s_i(t+1)=s_i(t)+  \alpha{B}_t h_i^+(t) +\beta{D}_t h_i^-(t),
\end{align}
where $\alpha,\beta>0$ are two positive constants marking the  weight each node put on the positive and negative recommendations.

The role of $h_i^+(t)$ in (\ref{9}) is consistent with the classical DeGroot's social learning model \cite{degroot} along trustful interactions.   In view of the definition of $h_i^-(t)$, in contrast to $h_i^+(t)$, the model is referred to as the state-flipping model.

{
\begin{remark}
The state-flipping model can be  interpreted as a situation where the neighbors connected by a negative link provide false values of their states to each node by flipping their true sign~\cite{altafini13}. Under this interpretation it is the {\it head} node along each negative arc that knows the sign of that arc. However, the tail node does not see the sign of the arc associated with the recommendations it receive.    The weights  and attentions  of recommendations, represented by $\alpha/\beta$ and $B_t/D_t$, respectively, are then descriptions of each node's possible  prior knowledge of the signs of its  neighbors.
\end{remark}

\begin{remark}
In standard consensus algorithms, nodes  communicate relative states. In other words, nodes hold  no absolute state information. For the state-flipping model to make sense, there must exist a global {\em origin} (state equal to 0) known by each node so that sign flipping is possible in the negative interactions.
\end{remark}
}

Let  $s(t)=\big(s_1(t)\dots s_n(t)\big)^T$ be the random vector representing the network state  at time $t$. The main objective of this paper is to analyze the behavior of the stochastic process $\{s(t)\}_{t\ge 0}$. In the following, we denote by $\mathds{P}$ the probability measure capturing all random components driving the evolution of the network state.

\section{Main Results}
In this section, we present our main results. We begin by stating two natural assumptions on the way nodes are selected for updates, and on the graph dynamics. In the first assumption, we impose that  at time $t$, any arc is selected with positive probability. The second assumption states that the unions of the graphs ${\cal G}_t$ over time-windows of fixed duration are strongly connected.

\noindent {\bf A1.} There is a constant  $p_\ast \in (0,1)$ such that for all $t\geq0$ and  $i,j\in \mathcal{V}$, $\mathds{P}\big( (i,j)\in {E}_t \big)\geq p_\ast$ if $(i,j)\in \mathcal{E}_t$.

\noindent {\bf A2.} There is an integer $K \geq 1$ such that the union graph $\mathcal{G}\big({[t,t+K-1]}\big)=\big(\mathcal{V}, \bigcup_{\tau \in [t,t+K-1] } \mathcal{E}_\tau\big)$ is strongly connected for all $t\geq0$.

The following theorem provides conditions under which the system dynamics converges almost surely. Surprisingly, these conditions are mild: we just require that the sum of the updating parameters $\alpha$ and $\beta$ is small enough, and that node updates occur with constant probabilities, i.e., $\mathds{E}\{B_t\}$ and $\mathds{E}\{D_t\}$ do not evolve over time. In particular, the state of each node converges almost surely even if the signs of the arcs change over time.

\medskip
\begin{thm}\label{thmsr1}
Assume that A1 and A2 hold, and that $\alpha,\beta>0$ are such that $\alpha+\beta  <{1}/{(n-1)}$. Further assume that for any $t\ge 0$, $b_t= b$ and $d_t= d$ for some $b,d \in (0,1)$. Then under the state-flipping  model, we have, for all $i\in\mathcal{V}$ and all initial state $s(0)$,
$
\mathds{P} \big( \lim_{t\rightarrow \infty} s_i(t) \ {\rm exists}\big)=1.
$
\end{thm}

\medskip
{
In the above theorem, we say that $\lim_{t\rightarrow \infty} s_i(t)$ exists if $s_i(t)$ converges to a finite limit as $t$ tends to infinity.

\begin{remark}
Theorem~\ref{thmsr1} shows an interesting property of  the state-flipping model: negative updates, together with the positive updates, contributes to the convergence of the node states whenever it holds that $\alpha+\beta  <{1}/{(n-1)}$. The condition $\alpha+\beta  \leq {1}/{(n-1)}$ guarantees that the absolute values of the node states are non-expansive for all signed graphs, compared to the state non-expansiveness of standard consensus algorithms \cite{mor}.
\end{remark}
}

Characterizing the limiting states is in general challenging. There are however scenarios where this can be done, which require the notion of structural balance~\cite{{harary56}}.

\begin{defn}
Let $\mathcal {G}=(\mathcal{V}, \mathcal{E})$ be a signed digraph. $\mathcal {G}$ is strongly balanced if  we can divide $\mathcal{V}$ into two disjoint  nonempty subsets $\mathcal{V}_1$ and $\mathcal{V}_2$ where negative arcs exist only between these two subsets.
\end{defn}

To predict the limiting system behavior, we make the following assumption.

\noindent
{\bf A3.}  $\{\mathcal{G}_t\}_{t\ge 0}$ is sign consistent.

Recall that  $\mathcal{G}_\ast$ denotes the total graph. The following theorem holds.

\medskip
\begin{thm}\label{thmsr2}
Assume that A1, A2 and A3 hold, and that $\alpha,\beta>0$ are such that $\alpha+\beta  <{1}/{(n-1)}$. Suppose $\mathcal{G}_\ast$ contains at least one negative arc and that every negative arc in $\mathcal{G}_\ast$ appears infinitely often in $\{\mathcal{G}_t\}_{t\ge 0}$. Further assume that for any $t\ge 0$, $b_t= b$ and $d_t= d$ for some $b,d \in (0,1)$. Then under the state-flipping model, we have, for any initial state $s(0)$:

(i) If $\mathcal{G}_\ast$ is strongly balanced, then there is a random variable $y_\ast$, with $y_\ast \leq \|s(0)\|_1$ almost surely, such that
$
\mathds{P} \big( \lim_{t\rightarrow \infty} s_i(t) = y_\ast, \forall i\in \mathcal{V}_1; \  \lim_{t\rightarrow \infty} s_i(t) =- y_\ast, \forall i\in \mathcal{V}_2\big)=1;
$

(ii) If $\mathcal{G}_\ast$ is not strongly balanced, then
 $
\mathds{P} \big( \lim_{t\rightarrow \infty} s_i(t) = 0, \forall i\in \mathcal{V}\big)=1.
$
\end{thm}

\medskip
Theorem \ref{thmsr2} states that strong structural balance is crucial to ensure convergence to nontrivial clustering states, which is consistent with the result of \cite{altafini13} derived for fixed graphs under continuous-time node updates. To establish the result, we do not rely on a spectral analysis as in \cite{altafini13}, but rather study the asymptotic behavior for each sample path. From the above theorem, we know that under the strong structural balance condition, the states of nodes in the same positive cluster converge to the same limit, and that the limits of two nodes in different positive clusters are exactly opposite. Using similar arguments as in \cite{altafini13}, the value of $y_\ast$ can be described as the limit of a random consensus process with the help of a gauge transformation.

Next we are interested in determining whether the states  could diverge depending on the values of the updating parameters $\alpha$ and $\beta$. We show that by increasing $\beta$, i.e., the strength of the negative recommendations,  one may observe such divergence. To this aim, we make the following assumptions.

\noindent {\bf A4.}  There is an integer $K\geq 1$ such that the union graph $\mathcal{G}^+\big({[t,t+K]}\big)=\big(\mathcal{V}, \bigcup_{\tau \in [t,t+K-1] } \mathcal{E}_\tau^+\big)$ is strongly connected for all $t\geq0$.

\noindent {\bf A5.}  There is an integer $K\geq 1$ such that the union graph $\mathcal{G}^-\big({[t,t+K]}\big)=\big(\mathcal{V}, \bigcup_{\tau \in [t,t+K-1] } \mathcal{E}_\tau^-\big)$ is strongly connected for all $t\geq0$.

 \noindent {\bf A6.} The events  $\{ (i,j)\in E_t \}$, $i,j\in\mathcal{V}$, $t=0,1,\dots$ are independent and there is a constant  $p^\ast \in (0,1)$ such that for all $t\geq0$ and  $i,j\in \mathcal{V}$, $\mathds{P}\big( (i,j)\in E_t \big)\leq   p^\ast$ whenever  $(i,j)\in \mathcal{E}_t$.

\medskip
\begin{prop}\label{thmsr3}
Assume that A1, A4, A5 and A6 hold, and that for any $t\ge 0$, $b_t= b$ and $d_t= d$ for some $b,d \in (0,1)$. Fix $\alpha \in[0,(4n)^{-1}]$. Then under the state-flipping  model, there is $\beta_\star >0$ such that whenever $\beta>\beta_\star$, we have
$
\mathds{P} \big( \lim_{t\rightarrow \infty} \max_{i\in \mathcal{V}} | s_i(t)| = \infty\big)=1
$
for almost all initial states $s(0)$.
\end{prop}
\medskip

{Proposition \ref{thmsr3} shows that under appropriate conditions, $\max_{i\in \mathcal{V}} | s_i(t)|$ diverges almost surely if the negative updating parameter $\beta$ is sufficiently large. We can in fact derive an explicit value for  $\beta_\star$.

\begin{remark}
The main difficulties of establishing Proposition  \ref{thmsr3} lie in the fact that we need on one hand to establish an absolute bound for the way $ \max_{i\in \mathcal{V}} | s_i(t)|$ decreases (which is obtained by a constructive proof), and on the other hand to establish a probabilistic lower bound for the possible increase of $ \max_{i\in \mathcal{V}} | s_i(t)|$ (which is obtained combining A4--A6  and by constructing and analyzing sample paths). These constructive derivations are rather conservative since we consider general random graph processes, but they nevertheless establish a positive drift for $\{ \max_{i\in \mathcal{V}} | s_i(t)|\}$ with an explicit  $\beta_\star$ so that almost sure divergence  is guaranteed.
\end{remark}

\begin{remark}
We also remark from Proposition  \ref{thmsr3} that a large deterministic weight on negative recommendations   leads to the divergence of the node states. It can also been seen from the forthcoming Lemma \ref{lem1} that if these weights on the recommendations are sufficiently small, $\max_{i\in \mathcal{V}} | s_i(t)|$ always converges no matter how the random  attentions $\{B_t\}$ and $\{D_t\}$ are selected.
\end{remark}
 }
Actually, one may even prove that when $\max_{i\in \mathcal{V}} | s_i(t)|$ grows large when $t\to\infty$, the state of  any node diverges. This result is referred to as the {\it no-survivor} property, and is formally stated in the following result.

\medskip
\begin{thm}\label{prop1}
Assume that A1, A2 and A6 hold, and that  for any $t\ge 0$, $b_t\equiv b$ and $d_t\equiv d$ for some $b,d \in (0,1)$. Fix the initial state $s(0)$. Then under the state-flipping  model, we have
$$
\mathds{P} \Big( \limsup_{t\rightarrow \infty} | s_i(t)| = \infty, i\in\mathcal{V} \ \Big| \limsup_{t\rightarrow \infty} \max_{i\in \mathcal{V}} | s_i(t)| = \infty\Big)=1.
$$
\end{thm}

\medskip

{

\begin{remark}
A similar kind of no-survivor property was first established in \cite{shiOR} under the model of repulsive  negative dynamics for pairwise node interactions. Theorem  \ref{prop1} establishes the same property for the considered state-flipping model, but for general random graph process. From the proof of Theorem  \ref{prop1}, it is clear that it is the arc-independence (assumption A6, see \cite{shiIT}), rather than synchronous or asynchronous node interactions, that directly results in the no-survivor divergence property for dynamics over signed random networks.
\end{remark}
}

In all above results, it can be seen from their proofs that extensions to time-varying  $\{b_t\}_{\geq 0}$ and $\{d_t\}_{\geq 0}$ are straightforward under mild assumptions. The resulting expressions are however more involved. We omit those discussions here to shorten the presentation.

\section{Proofs}
In this section, we present the detailed proofs of the results stated in the previous section. We first establish some technical lemmas, and then the proofs of each result.
\subsection{Supporting Lemmas}

For any $t\ge 0$, we define $M(t)=\max_{i\in \mathcal{V}} |s_i(t)|$ and  $Y_i(t)=\alpha B_t|N_i^+(t)| +\beta  D_t|N_i^-(t)|$, which will be used throughout the whole paper.

\medskip
{
\begin{lem}\label{lem1}
 Suppose $\alpha+\beta \leq {1}/{(n-1)}$. Then $M(t+1)\le M(t)$.
\end{lem}
\noindent{\it Proof.} Observe that  $|N_i^+(t)|+|N_i^-(t)|\leq n-1$. Hence $Y_i(t)\in [0,1]$ as long as $\alpha+\beta \leq {1}/{(n-1)}$. Now for any $i\in {\cal V}$,
\begin{align}
|s_i(t+1)|
&=\Big|s_i(t)-  \alpha{B}_t \sum_{j\in{N}_i^+(t) } \big(s_i(t)-s_j(t)\big) \nonumber\\
 &\ -\beta{D}_t \sum_{j\in{N}_i^-(t) } \big(s_i(t)+s_j(t)\big)\Big| \nonumber\\
&=\Big|\big(1-Y_i(t)\big)s_i(t)+  \alpha{B}_t \sum_{j\in{N}_i^+(t) }s_j(t) \nonumber\\
&\  -\beta{D}_t \sum_{j\in{N}_i^-(t) } s_j(t) \Big| \nonumber\\
&\leq \Big( \big| 1-Y_i(t) \big| +Y_i(t) \Big)  \max_{j\in \mathcal{V}} |s_j(t)|\nonumber\\
&= \max_{j\in \mathcal{V}} |s_j(t)|, \nonumber
\end{align}
which completes the proof. \hfill$\blacksquare$

\begin{remark}
Lemma \ref{lem1} establishes the non-expansiveness property of the considered  model.  It's clear  from its proof that the condition $\alpha+\beta \leq {1}/{(n-1)}$ in Lemma \ref{lem1} can be relaxed to $\alpha+\beta\leq 1/ {\rm Deg}(\mathcal{G}_t)$, where ${\rm Deg}(\mathcal{G}_t)$ denotes the maximum degree of the graph $\mathcal{G}_t$. Here for convenience we use the current statement since ${\rm Deg}(\mathcal{G}_t)\leq n-1$ for all $t$.
\end{remark}

}

\medskip
\begin{lem}\label{lem2}
Assume that $\alpha+\beta \leq {1}/{(n-1)}$. Let $i\in\mathcal{V}$ and assume that $|s_i(t)|\leq  \zeta _0 M(t)$ for some $0<\zeta_0<1$. Then
$$
 |s_i(t+ k)|\leq  \big(1-  (1-\zeta_0)\gamma_\ast^k \big) M(t), \quad k=0,1,\dots
$$
 where $\gamma_\ast=1- (\alpha+\beta)(n-1)$.
\end{lem}
\noindent{\it Proof.}  We have:
\begin{align}
&|s_i(t+1)|\leq \Big( 1-Y_i(t) \Big) |s_i(t)|  + Y_i(t) M(t)\nonumber\\
&\leq \Big( 1-Y_i(t) \Big) \zeta_0M(t)  + Y_i(t) M(t)\nonumber\\
&\leq  \Big(1-(\alpha+\beta)(n-1) \Big) \zeta_0 M(t)  +(\alpha+\beta)(n-1) M(t) \nonumber\\
&=\big(1-  (1-\zeta_0)\gamma_\ast \big) M(t). \nonumber
\end{align}
The lemma is then obtained by applying a simple induction argument. \hfill$\blacksquare$

\medskip
\begin{lem}\label{lem3}
Assume that $\alpha+\beta \leq {1}/{(n-1)}$. Let $i\in\mathcal{V}$ and assume that $|s_i(t)|\leq  \zeta _0 M(t)$ for some $0<\zeta_0<1$. Let $(i,j)\in\mathcal{E}_t$. Then conditioned on
 $B_t=1$ if $(i,j) \in \mathcal{E}_t^+$, $D_t=1$ if $(i,j) \in \mathcal{E}_t^-$, we have
$$
|s_j(t+ 1)| \leq  \big(1-  (1-\zeta_0)\min\{\alpha,\beta\}\big) M(t).
$$
\end{lem}{
\noindent{\it Proof.} Suppose $(i,j) \in \mathcal{E}_t^+$ with $B_t=1$. Then we have
\begin{align}\label{r3}
|s_j(t+1)|
&=\Big|s_j(t)-  \alpha{B}_t \sum_{k\in{N}_j^+(t) } \big(s_j(t)-s_k(t)\big) \nonumber\\
&\ -\beta{D}_t \sum_{k\in{N}_j^-(t) } \big(s_j(t)+s_k(t)\big)\Big| \nonumber\\
&=\Big|\alpha s_i(t)+\big(1-Y_j(t)\big)s_j(t) \nonumber\\
&\  + \alpha{B}_t \sum_{k\in{N}_j^+(t)\setminus\{i\} } s_k(t)-\beta{D}_t \sum_{k\in{N}_j^-(t) }s_k(t)\Big| \nonumber\\
&\leq \alpha|s_i(t)| +\Big(1-Y_j(t) +Y_j(t) -\alpha \Big)M(t) \nonumber\\&\leq  \min\{\alpha,\beta\}  |s_i(t)|  + \big(1- \min\{\alpha,\beta\} \big) M(t),
\end{align}
where in the last inequality we have used the fact that $|s_i(t)|\leq M(t)$. It is straightforward to see that (\ref{r3}) continues to hold with $D_t=1$ if $(i,j) \in \mathcal{E}_t^-$.
Plugging in the assumption that $|s_i(t)|\leq  \zeta _0 M(t)$ into (\ref{r3}), one gets the desired inequality. This proves the lemma.  \hfill$\blacksquare$
}

Note that if the conditions in Lemmas \ref{lem2} and \ref{lem3} are replaced by $|s_i(t)|<  \zeta _0 M(t)$, then we have the same conclusions but with strict inequalities. Moreover, in view of Lemma \ref{lem1}, the following limit is well defined: $M_\ast= \lim_{t\rightarrow \infty} M(t)$.

\medskip
\begin{lem}\label{lem4}
Assume that A1 and A2 hold, $\alpha,\beta>0$, and  $\alpha+\beta  < {1}/{(n-1)}$. Further assume that for any $t\ge 0$, $b_t\equiv b$ and $d_t\equiv d$ for some $b,d \in (0,1)$. Then for any initial state $s(0)$, we have $\mathds{P} \big( \lim_{t\rightarrow \infty} |s_i(t)|=M_\ast, \forall i\in\mathcal{V}\big)=1$.
\end{lem}

\noindent{\it Proof.} We prove this lemma using sample path arguments by contradiction. Let us assume that:

 \noindent {H1.} There exist $i_0\in\mathcal{V}$ and  $\delta, q_\ast\in (0,1)$ such that
$
\mathds{P} \big( \liminf_{t\rightarrow \infty} |s_{i_0}(t)|<\delta M_\ast\big)\geq q_\ast.
$

Let $\epsilon >0$. Define
$$
T(\epsilon):=\inf_{k\geq 0} \big\{ M(t)\leq (1+\epsilon) M_\ast, \forall t\geq k \big\}
$$
and
$$
T^\ast:=\inf_{t\geq T(\epsilon)} \big\{s_{i_0}(t) <\delta M_\ast \big\}.
$$
Note that $T(\epsilon)$ is a stopping time, and the monotonicity of $M(t)$ guarantees that $T$ is bounded almost surely~\cite{durr}. Moreover, $T^\ast$ is also a stopping time, and it is bounded with probability at least $q_\ast$ in view of H1. Next, we use Lemmas~\ref{lem2} and \ref{lem3} to get a contradiction. {  Plugging in the fact that $M(T^\ast)\leq  M_\ast(1+\epsilon)$ and invoking  Lemma~\ref{lem2},  conditioned on $\{T^\ast <\infty\}$, we have that for all $k=0,1,\dots$:
\begin{align}
\big|s_{i_0}(T^\ast+k)\big| <\big(1- (1-\delta)\gamma_\ast^k \big)  M_\ast(1+\epsilon).
\end{align}
}

Now  consider the time interval $[T^\ast, T^\ast+K-1]$.  The independence of  $\{{B}_t\}_{t\ge 0}$, $\{{D}_t\}_{t\ge 0}$, and $\{G_t\}_{t\ge 0}$  guarantees that
 $({G}_{T^\ast}, {B}_{T^\ast}, {D}_{T^\ast}), ( {G}_{T^\ast+1}, {B}_{T^\ast+1}, {D}_{T^\ast+1}),\dots$ are independent random variables, and they are independent of $\mathcal{F}_{{T^\ast}-1}$ (cf., Theorem 4.1.3 in \cite{durr}). From their definitions we also know that $({B}_{T^\ast}, {D}_{T^\ast})$,  $({B}_{T^\ast+1}, {D}_{T^\ast+1})$, $\dots$ are i.i.d. with the same distribution as $({B}_{0}, {D}_{0})$, and Assumption A2 guarantees that  $\mathcal{G}\big({[T^\ast,T^\ast+K-1]}\big)=\big(\mathcal{V}, \bigcup_{\tau \in [T^\ast,T^\ast+K-1] } \mathcal{E}_\tau\big)$ is strongly connected. Therefore, there exists a node $i_1\neq i_0$ and $\tau_1\leq K$ such that $(i_0,i_1) \in \mathcal{E}_{T^\ast+\tau_1}$ (note that $i_1$ and $\tau_1$ are random variables, but they are independent with $\mathcal{F}_{{T^\ast}-1}$ since $T^\ast$ is a stopping time). Hence we can apply Lemma \ref{lem3} and conclude that
\begin{align}
\big|s_{i_1}(T^\ast +\tau_1)\big| < \big(1-(1-\delta)\gamma_\ast^{\tau_1}\min\{\alpha,\beta\}\big)M_\ast(1+\epsilon) \nonumber
\end{align}
with a probability at least $p_\ast\min\{b,d\}$.  Taking $\zeta_0=1-(1-\delta)\gamma_\ast^{\tau_1}\min\{\alpha,\beta\}$ for the $\zeta_0$ introduced in Lemma \ref{lem2}, we have
$$
1-  (1-\zeta_0)\gamma_\ast^k  =1-(1-\delta)\gamma_\ast^{\tau_1+k}\min\{\alpha,\beta\}.
$$
Therefore, applying Lemma \ref{lem2} (note that we can replace $M(t)$ with $M_\ast(1+\epsilon)$  in Lemma \ref{lem2}) we have that for all $k=K,K+1,\dots$,
\begin{align}
\big|s_{i_1}(T^\ast +k)\big| < \big(1-(1-\delta)\gamma_\ast^{k}\min\{\alpha,\beta\}\big)M_\ast(1+\epsilon). \nonumber
\end{align}

We can repeat the same argument over time intervals $[T^\ast+K,T^\ast+2K-1],\dots,[T^\ast+(n-2)K,T^\ast+(n-1)K-1]$. {Assuming that the node set $I_{k}:=\{i_0,\dots,i_{k-1}\}$ is selected, it follows from the strong connectivity assumption A2 that there exists an arc from $I_k$ to $\mathcal{V}\setminus I_k$ in the union graph of the corresponding interval. In this way we add the tail node of such arc into $I_k$ and obtain $I_{k+1}$ for $k=1,\dots,n-1$. } We can thus recursively  find $i_2,\dots,i_{n-1}$ with $\mathcal{V}=\{i_0,\dots,i_{n-1}\}$ and bound the absolute values of their states. Finally, we get:
\begin{align}\label{1}
&\mathds{P} \big(M(T^\ast +(n-1)K) <  \big[1-\gamma_\ast^{(n-1)K}  (\min\{\alpha,\beta\})^{n-1} \nonumber\\
& \times(1-\delta)\big]  M_\ast(1+\epsilon) \big| T^\ast <\infty\big)\geq \big (p_\ast \min\{b,d\}\big)^{n-1}.
\end{align}
Now select $\epsilon$ sufficiently small so that $ \theta:=\big(1-(1-\delta)\gamma_\ast^{(n-1)K}(\min\{\alpha,\beta\})^{n-2}\big)(1+\epsilon)<1$. Using the monotonicity of $M(t)$ established in Lemma \ref{lem1}, we deduce from (\ref{1}):
\begin{align}
\mathds{P} \big(M_\ast < \theta M_\ast \big| T^\ast<\infty\big)\geq \big (p_\ast\min\{b_\ast,d_\ast\}\big)^{n-1}, \nonumber
\end{align}
which is impossible and hence, H1 is not true. We have proved that:
$$
\mathds{P} \big( \liminf_{t\rightarrow \infty} |s_{i}(t)|= M_\ast, \forall i\in\mathcal{V}\big)=1.
$$
The claim then follows easily from Lemma \ref{lem1}. \hfill$\blacksquare$

{
\begin{remark}
It is easy to see from the proof  that Lemma~\ref{lem4} continues to hold if we relax the requirement of $b_t,d_t$ to $0<\underline{b}\leq b_t \leq 1$ and $0<\underline{d}\leq d_t\leq 1$ for some $\underline{b}, \underline{d}\in(0,1)$. Lemma~\ref{lem4} indicates that with sufficient connectivity on the graphs defining the dynamical   environment (Assumption A2), the absolute values of the nodes states, will eventually converge to a consensus with probability one under quite general conditions on how the random interactions take place in the environment. Noting that A2 is imposed on the overall underlying graph, this concludes that both the positive and negative links contribute to the node states' consensus in absolute value.
\end{remark}
}

\begin{lem}\label{lem5}
Let $\alpha<(4n)^{-1}$ and $\beta>16n^{n-1}$. Then $M(t+1)\geq (2n)^{-1} M(t) $ defines a sure event.
\end{lem}
\noindent{\it Proof.} Let us first assume that $D_t=0$. Let $i\in\mathcal{V}$ such that $|s_i(t)|=M(t)$. Then with $\alpha<(4n)^{-1}$, we have
\begin{align}
&M(t+1)\geq|s_i(t+1)|\nonumber\\
&\geq \big|1- \alpha B_t |N_i^{+}(t)| \big| \cdot|s_i(t)|-\alpha B_t |N_i^{+}(t)|\cdot M(t)\nonumber\\
&\geq \big|1- 2\alpha B_t |N_i^{+}(t)| \big| \cdot M(t) \nonumber\\
&\geq {1\over n} M(t). \nonumber
\end{align}

Now assume that $D_t=1$. We first prove the following claim.

\noindent
{\it Claim.} Suppose there exits  $i_1 \in \mathcal{V}$ such that $|s_{i_1}(t)| \in\big[(1-Z_2  )M(t), (1-Z_1  )M(t)\big]$ with $0\leq Z_1  <Z_2  <nZ_2  <1/4$ and $\beta Z_2  \geq 2$. Then $\mathcal{H}_1\bigcup \mathcal{H}_2$ is a sure event, where
$$
\mathcal{H}_1=\big\{M(t+1)\geq M(t)/4\big\}
$$
and
$$
\mathcal{H}_2=\Big\{\exists i_2 : |s_{i_2} (t)|\in \big[(1-nZ_2  )M(t), (1-Z_2  )M(t)\big]\Big \}.
$$
To prove this claim, we distinguish three cases:\\
\begin{itemize}
\item [(i)] Let $s_{i_1}(t) \in\big[(1-Z_2  )M(t), (1-Z_1  )M(t)\big]$ and assume that there exists $j_\ast\in\mathcal{V}$ such that $j_\ast\in N^-_{i_1}(t)$ and $s_{j_\ast}(t)\in \big[-(1-nZ_2  ) M(t), M(t)\big]$.
Then $s_{i_1}(t)+s_{j_\ast}(t)\geq (n-1)Z_2  M(t)\geq0$ and $s_{i_1}(t)+s_{j}(t)\geq -Z_2  M(t)$ for all $j\in \mathcal{V}\setminus\{i_1,j_\ast\}$. Thus, taking out the term $s_{i_1}(t)+s_{j_\ast}(t)$ in $h_{i_1}^-(t)$ from (\ref{9}), some simple algebra leads to 
\begin{align}\label{6}
&M(t+1)\geq |s_{i_1}(t+1)| \nonumber\\
 &\geq  \beta\big|s_{i_1}(t)+s_{j_\ast}(t)\big|-M(t)- 2\alpha(n-1)M(t)\nonumber\\
 &\ \ -\beta (n-2)Z_2 M(t) \nonumber\\
&\geq \big| \beta Z_2  -1 - ({n-1})({2n})^{-1}\big|\cdot M(t)\nonumber\\
&\geq \frac{1}{2} M(t),
\end{align}
 where in the last inequality we have used the assumption that $\beta Z_2  \geq 2$.

\item [(ii)]  Let $s_{i_1}(t) \in\big[(1-Z_2  )M(t), (1-Z_1  )M(t)\big]$ and assume that $s_{j}(t)\in[-M(t),-(1-Z_1)(M(t))]$ for all  $j\in N^-_{i_1}(t)$, and, more generally, $s_{i_1}(t)+s_{j}(t)\leq 0$ for all $j\in N^-_{i_1}(t)$, which implies that $h_{i_1}^{-}(t)\geq 0$.  Observing that $s_{i_1}(t)\geq 0$, we obtain 
\begin{align}\label{7}
&M(t+1)\geq |s_{i_1}(t+1)| \nonumber\\
 &\geq |s_{i_1} (t)|- 2\alpha(n-1)M(t)\nonumber\\
&\geq \big| 1- Z_2   - ({n-1})({2n})^{-1}\big|\cdot M(t)\nonumber\\
&\geq \frac{1}{4} M(t).
\end{align} 

\item[(iii)] Let $m_1,\dots,m_\ell \in \mathcal{V}$ and $w_1,\dots, w_\jmath$ satisfy
$$
s_{m_\varrho} (t) \in \big[(1-Z_2  )M(t), (1-Z_1  )M(t)\big],  \ \ \varrho=1,\dots,\ell,
$$
and
$$
s_{w_\varrho} (t) \in \big[-(1-Z_1  )M(t), -(1-Z_2  )M(t)\big],  \ \ \varrho=1,\dots,\jmath,
$$
respectively. Without loss of generality we assume the existence of such ${m_\varrho}$ and ${w_\varrho}$ since otherwise the desired conclusions immediately falls to Case (i) and (ii).

Now  without loss of generality  suppose $s_{m_1}(t)=\min_{\varrho=1,\dots,\ell} s_{m_\varrho} (t)$. From Case (i) and (ii), the desired claim can possibly be violated only when there exists ${w_{\varrho_\ast}}\in N^-_{m_1}(t)$ with $s_{w_{\varrho_\ast}}(t)+s_{m_1} (t) > 0$ for some $\varrho_\ast\in\{1,\dots,\jmath\}$ (otherwise we can bound $|s_{m_1}(t+1)|$ from Case (ii)). While due to the choice of $m_1$ it holds that
$$
s_{w_{\varrho_\ast}}(t)+s_{m_\varrho} (t) > 0,\ \ \varrho=1,\dots,\ell.
$$
We can therefore denote $y_i(t)=-s_i(t),i\in\mathcal{V}$ and obtain that $y_{w_{\varrho_\ast}}(t) \in\big[(1-Z_2  )M(t), (1-Z_1  )M(t)\big]$ and $y_{w_{\varrho_\ast}}(t)+y_{m_\varrho} (t) <0,\ \ \varrho=1,\dots,\ell.$ We can thus establish  bound for $|y_{w_{\varrho_\ast}}(t)|=|s_{w_{\varrho_\ast}}(t)|$, again applying Case (ii).
\end{itemize}

From the above three cases, we deduce that if $\mathcal{H}_2$ does not hold, then $\mathcal{H}_1$ must be true. This proves the claim.

Finally,  we complete the proof of the lemma using the claim we just established. Take $\epsilon= 8^{-1} n^{-n-1}$ and $\beta=16 n^{n+1}$. We proceeds in steps.

\vspace{2mm}

\noindent S1) Let $m_1 \in \mathcal{V}$ with $|s_{m_1}(t)|=M(t)$.  Applying the claim with $Z_1=0$ and $Z_2=\epsilon$, we deduce that either the lemma holds or there is another node $m_2 \in \mathcal{V}$ such that  $|s_{m_2}(t)|\in \big[(1-n \epsilon )M(t), (1-\epsilon )M(t)\big]$.

\vspace{2mm}

\noindent S2) If in the first step, we could not conclude that the lemma holds, we can apply the claim to $m_2$ and then obtain that either the lemma holds, or there is a node $m_3$ such that  $|s_{m_3}(t)|\in \big[ (1-n^2 \epsilon )M(t),  (1-n \epsilon )M(t)\big]$.

\vspace{2mm}

The argument can be repeated for $m_3,\dots$ applying the claim adapting the value of $\epsilon$ and $\beta$. Since there are a total of $n$ nodes, the above repeated procedure necessarily ends, so the lemma holds. \hfill$\blacksquare$

{

\begin{remark}
The purpose of Lemma \ref{lem5} is to establish an absolute lower bound regarding the possible decreasing of $M(t)$. This lower bound is absolute in the sense that it does not depend on  the random graph processes, and require the constructive conditions  $\alpha<(4n)^{-1}$ and $\beta>16n^{n+1}$ to hold. These conditions are certainly rather conservative for a particular  node interaction process, e.g.,  the pairwise gossiping model \cite{boyd}, or the  i.i.d. link failure model \cite{kar2}.
\end{remark}

}

\subsection{Proof of Theorem \ref{thmsr1}}
From Lemma \ref{lem4}, we know that for any $i\in\mathcal{V}$, one of the following events happens almost surely:
$\big\{ \lim_{t\rightarrow \infty} s_{i}(t)= M_\ast\}$; $\big\{ \lim_{t\rightarrow \infty} s_{i}(t)=-M_\ast\}$; $\big\{ \liminf_{t\rightarrow \infty} s_{i}(t)= -M_\ast$ and $ \limsup_{t\rightarrow \infty} s_{i}(t)= M_\ast\}$. Therefore, we just need to rule out the last case. We actually prove that:
 $
 \mathds{P}\big(M_\ast>0, \liminf_{t\rightarrow \infty} s_{i}(t)= -M_\ast, \limsup_{t\rightarrow \infty} s_{i}(t)= M_\ast, \lim_{t\rightarrow \infty} |s_{i}(t)|= M_\ast\big)=0.
 $

The following claim holds.

\vspace{2mm}
{  \noindent {\it Claim}. Suppose $\alpha+\beta \leq {1}/{(n-1)}$. Then $s_i(t+k)\leq \gamma_\ast^k s_i(t)+ (1-\gamma_\ast^k) M(t)$ for all $k\geq 0$, where $\gamma_\ast=1- (\alpha+\beta)(n-1)$ was introduced in Lemma~\ref{lem2}.

\vspace{2mm}
First we note that
\begin{align}
s_i(t+1)
&=\big(1-Y_i(t)\big)s_i(t)+  \alpha{B}_t \sum_{j\in{N}_i^+(t) }s_j(t)\nonumber\\
&\  -\beta{D}_t \sum_{j\in{N}_i^-(t) } s_j(t) \nonumber\\
&\leq \big(1-Y_i(t)\big)s_i(t)  +Y_i(t) M(t)\nonumber\\
&\leq  \gamma_\ast s_i(t)+ (1-\gamma_\ast) M(t), \nonumber
\end{align}
where the last inequality holds from the facts that $1-Y_i(t)\geq 1-(\alpha+\beta)(n-1) = \gamma_\ast $ and that $s_i(t)\leq M(t)$.  A simple recursive analysis leads to the claim immediately.

Now take $\epsilon>0 $ and define $T_1(\epsilon):=\inf\big\{ k : M(k)\leq M_\ast(1+\epsilon)\big\}$. Note that $T_1(\epsilon)$ is a stopping time due to the monotonicity of $M(t)$ established in Lemma~\ref{lem1}. In light of the above claim  we get
\begin{align}\label{2}
s_i(t+k)&\leq \gamma_\ast^k s_i(t)+ (1-\gamma_\ast^k)M_\ast(1+\epsilon )
\end{align}
for all $k=0,1,\dots$ and $t\geq T_1$.
}

Let $M_\ast >0$. Assume that $\liminf_{t\rightarrow \infty} s_{i}(t)= -M_\ast$. Then for the given $\epsilon$, we can find an infinite sequence $T_1(\epsilon)<t_1<t_2<\dots$ such that $s_{i}(t_m) \leq -M_\ast(1-\epsilon)$. Now, if $\limsup_{t\rightarrow \infty} s_{i}(t)= M_\ast$, for any $t_m$, we can find $\bar{t}_m>t_m$ with $s_{i}(\bar{t}_m)\geq  M_\ast(1-\epsilon)$. Then based on (\ref{2}), there must be $\hat{t}_m \in[t_m,\bar{t}_m]$ such that (see Fig. \ref{proof})
{
\begin{align}
&-\gamma_\ast M_\ast(1-\epsilon)+ (1-\gamma_\ast)M_\ast(1+\epsilon ) \leq s_i(\hat{t}_m) \nonumber\\
&\  \leq  -\gamma_\ast^2 M_\ast(1-\epsilon)+ (1-\gamma_\ast^2)M_\ast(1+\epsilon ).
\end{align}
We then deduce that for all $m=1,2,\dots$,
 \begin{align}\label{r1}
\big|s_i(\hat{t}_m) \big|&\leq \max \Big\{ \big| -\gamma_\ast M_\ast(1-\epsilon)+ (1-\gamma_\ast)M_\ast(1+\epsilon )\big|, \nonumber\\
 &\ \big|  -\gamma_\ast^2 M_\ast(1-\epsilon)+ (1-\gamma_\ast^2)M_\ast(1+\epsilon ) \big| \Big\} \nonumber\\
&=  M_\ast\max \Big\{ \big| 1-2\gamma_\ast+\epsilon\big|, \big| 1-2\gamma_\ast^2+\epsilon\big|  \Big\}\nonumber\\
&\leq M_\ast \Big(\epsilon+ \max \big\{ \big| 1-2\gamma_\ast\big|, \big| 1-2\gamma_\ast^2\big|  \big\}\Big)\nonumber\\
&\leq M_\ast\Big(1+ \max \big\{ \big| 1-2\gamma_\ast\big|, \big| 1-2\gamma_\ast^2\big|  \big\}\Big)/2
\end{align}
if  we choose $\epsilon$ sufficiently small so that $$
\epsilon<\Big(1- \max \big\{ \big| 1-2\gamma_\ast\big|, \big| 1-2\gamma_\ast^2\big|  \big\}\Big)/2.
 $$

\begin{figure}[t]
\begin{center}
\includegraphics[height=0.95in]{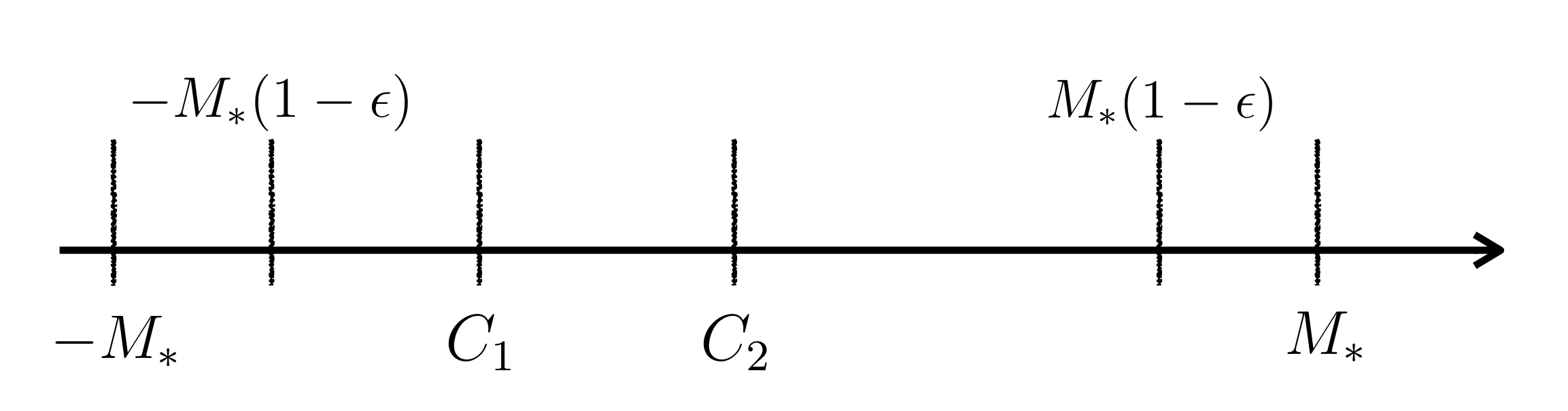}
\caption{An illustration of the existence  of $\hat{t}_m$ in the proof of  Theorem \ref{thmsr1}. Here $C_1:=-\gamma_\ast M_\ast(1-\epsilon)+ (1-\gamma_\ast)M_\ast(1+\epsilon )$, $C_2:=-\gamma_\ast^2 M_\ast(1-\epsilon)+ (1-\gamma_\ast^2)M_\ast(1+\epsilon )$. By  (\ref{2}), $s_i$ starting from the interval $[-M_\ast, -M_\ast(1-\epsilon)]$ entering $[M_\ast(1-\epsilon), \infty)$ must go through the interval $[C_1,C_2]$. } \label{proof}
\end{center}
\end{figure}

Now we see that (\ref{r1}) contradicts  $\lim_{t\rightarrow \infty} |s_{i}(t)|= M_\ast$ since
 $$
 0<\Big(1+ \max \big\{ \big| 1-2\gamma_\ast\big|, \big| 1-2\gamma_\ast^2\big|  \big\}\Big)/2<1
 $$ when $0<\gamma_\ast<1$. We have completed the proof  of Theorem \ref{thmsr1}.
 }
\subsection{Proof of Theorem \ref{thmsr2}} {  In view of Theorem \ref{thmsr1}, with probability one we can divide the node set $\mathcal{V}$ into the following two subsets of nodes
$$
\mathcal{V}_1^\ast:=\big \{i\in\mathcal{V}: \lim_{t \rightarrow \infty} s_i(t)=-M_\ast\big \}
 $$
 and
 $$\mathcal{V}_2^\ast:=\big \{i\in\mathcal{V}: \lim_{t \rightarrow \infty} s_i(t)=M_\ast\big \}.
$$
Apparently at least one of the sets is non-empty.
Without loss of generality, we assume $\mathcal{V}_1^\ast \neq \emptyset$ and $\mathds{P}(M_\ast>0)>0$ for the rest of the proof. With Assumption A3, each arc $(i,j)\in \mathcal{E}_\ast$ is associated with a unique sign. We therefore denote the sign of $(i,j)\in \mathcal{E}_\ast$ as $\sigma_{ij}^\ast$.  To establish the desired conclusion we first show the following claim holds.

\noindent{\it Claim.} If $(i,j)\in \mathcal{E}_\ast$ with $i,j\in \mathcal{V}_1^\ast$, then $\sigma_{ij}^\ast=+$.

 The above claim indicates that the arcs among nodes in $\mathcal{V}_1^\ast$ are necessarily positive. We now prove this claim using a similar sample-path analysis as the proof of Theorem \ref{thmsr1} by a contradiction argument. Suppose there exist $i^\dag,j^\dag\in \mathcal{V}_1^\ast$ such that $(i^\dag,j^\dag)\in \mathcal{E}_\ast$ with $\sigma_{i^\dag j^\dag}^\ast=-$. By our assumption the arc $(i^\dag,j^\dag)$ appears infinitely often in $\{\mathcal{G}_t\}_{t\geq 0}$. This means that there exists an infinite subsequence $\{t_m\}_{m=0}^\infty$ such that  $(i^\dag,j^\dag)\in \mathcal{E}_{t_m}$ for all $m\geq 0$. We assume with out loss of generality that $(i^\dag,j^\dag)\in \mathcal{E}_{t}$ for all $t\geq 0$ since the following analysis can  indeed be carried out along the subsequence $\{t_m\}_{m=0}^\infty$.

 Based on the definition of $\mathcal{V}_1^\ast$ and $\mathcal{V}_2^\ast$, for any $\epsilon>0$, we can define
\begin{align}
&T^\dag_1(\epsilon):=\inf\Big\{ k : s_i (k) \in  \big[-M_\ast(1+\epsilon), -M_\ast(1-\epsilon)\big],  i\in \mathcal{V}^\ast_1; \nonumber\\
& \ \ s_i (k) \in  \big[M_\ast(1-\epsilon), M_\ast(1+\epsilon)\big],  i\in \mathcal{V}^\ast_2 \Big\}.\nonumber
\end{align}
Since $(i^\dag,j^\dag)\in \mathcal{E}_{t}$ for all $t\geq 0$ with $\sigma_{i^\dag j^\dag}^\ast=-$, we conclude that
\begin{align}
s_{j^\dag}(T^\dag_1+1)
&=\big(1-Y_{j^\dag}(T^\dag_1)\big)s_{j^\dag}(T^\dag_1)- \beta{D}_{T^\dag_1} s_{i^\dag}(T^\dag_1) \nonumber\\
&\ + \alpha{B}_{T^\dag_1} \sum_{k\in{N}_{j^\dag}^+(T^\dag_1) }s_k(T^\dag_1) \nonumber\\
 &\ -\beta{D}_{T^\dag_1} \sum_{k\in{N}_{j^\dag}^-(t)\setminus\{i^\dag\} } s_k(T^\dag_1) \nonumber\\
&\geq -\big(1-Y_{j^\dag}(T^\dag_1)\big)M_\ast(1+\epsilon)+ \beta M_\ast(1-\epsilon)\nonumber\\
&\ -Y_{j^\dag}(T^\dag_1)M_\ast(1+\epsilon)\nonumber\\
&=-\Big[1-\beta+\epsilon(1+\beta)\Big]M_\ast \nonumber\\
&\geq -\frac{1-\beta}{2}M_\ast \nonumber\\
&>-(1-\epsilon)M_\ast
\end{align}
if ${D}_{T^\dag_1}=1$ and $\epsilon$ is chosen to satisfy
$\epsilon<(1-\beta)/(2(1+\beta))$.

We can recursively define
\begin{align}
&T^\dag_{m+1}(\epsilon)\nonumber\\
&\ \ :=\inf\Big\{ k\geq T^\dag_m: s_i (k) \in  \big[-M_\ast(1+\epsilon), -M_\ast(1-\epsilon)\big], \nonumber\\& \ \ \ \  i\in \mathcal{V}^\ast_1;
s_i (k) \in  \big[M_\ast(1-\epsilon), M_\ast(1+\epsilon)\big],  i\in \mathcal{V}^\ast_2 \Big\}.\nonumber
\end{align}
Repeating the above analysis   we have
 \begin{align}\label{r2}
s_{j^\dag}(T^\dag_m+1)>-(1-\epsilon)M_\ast
\end{align}
for each $m\geq 1$ conditioned on that  ${D}_{T^\dag_m}=1$. Note that $\big\{{D}_{T^\dag_m}\big\}_1^\infty$ defines a sequence of independent random variables since $\{T^\dag_m\}$ are a sequence of stopping times. We can therefore invoke the second Borel-Cantelli Lemma (e.g., Theorem 2.3.6 in \cite{durr}) to conclude that (\ref{r2}) holds for infinitely many $m$. In other words, we have established that
$$
\liminf_{t\rightarrow \infty} s_{j^\dag}(T^\dag_m+1)>-(1-\epsilon)M_\ast
$$
for any $\epsilon<(1-\beta)/(2(1+\beta))$. This contradicts the fact that   $j^\dag\in\mathcal{V}_1^\ast$. We have now proved the given claim.

Having established the above claim, the rest of the argument becomes straightforward. Next, the following analysis will be carried out for  the two cases in the theorem statement.

\noindent {(i).} The total graph $\mathcal{G}_\ast$ is strongly balanced with nonempty $\mathcal{V}_1$ and $\mathcal{V}_2$, and hence ${\cal V}_1^\ast$ is for example included in ${\cal V}_1$, which in turns implies that ${\cal V}_2^\ast \neq \emptyset$. Again there are only positive arcs among  nodes of ${\cal V}_2^\ast$. We simply deduce that $\{\mathcal{V}_1,\mathcal{V}_2\}= \{ \mathcal{V}_1^\ast,\mathcal{V}_2^\ast\}$. Thus the required $y_\ast$ is exactly $M_\ast$.

\noindent {(ii).} We take a contradiction argument. Since $\mathds{P}(M_\ast >0)>0$, we have $\mathcal{V}_1^\ast\cap \mathcal{V}_2^\ast=\emptyset$.  Again arcs between nodes in the same set from $\mathcal{V}_i^\ast, i=1,2$ are necessarily positive. However there is at least one negative link in $\mathcal{G}_\ast$ by  assumption, which can only be an arc between $\mathcal{V}_1^\ast$ and $\mathcal{V}_2^\ast$.   Thus both $\mathcal{V}_1^\ast$ and $\mathcal{V}_2^\ast$ are nonempty, which implies that $\mathcal{G}_\ast$ must be strongly balanced.

The proof is complete.

}

\subsection{Proof of Proposition  \ref{thmsr3}}
Let $\beta>16n^{n+1}$ so the conditions of Lemma \ref{lem5} hold. Let us fix $t\geq 0$ and assume that $|s_{i_0}(t)|=M(t)$ for some $i_0 \in \mathcal{V}$. By symmetry, we can also assume without loss of generality that $s_{i_0}(t)=M(t)$. Let $i_\ast \in \mathcal{V}\setminus\{i_0\}$. Under Assumptions A4 and A6, we prove the following claim.

\noindent{\it Claim.} There is an integer $N_0\geq 1$ and $q_0>0$ such that
$$
\mathds{P}\Big(s_{i_0}(t+N_0K)=M(t),s_{i_\ast}(t+N_0K) \geq {M(t)}/{2}\Big) \geq q_0.
$$

In view of the connectivity assumption A4 and the arc independence assumption A6, the event $$
\big\{s_{i_0}(t+N_0K)=M(t), s_{i_\ast}(t+N_0K-1) \geq {M(t)}/{2} \big\}
 $$given  $s_{i_0}(t)=M(t)$ can be easily constructed by selecting a proper sequence of positive arcs for time slots $t,t+1,\dots,t+N_0K-1$, and by imposing that $B_\tau=1,D_\tau=0,$ $\tau=t,t+1,\dots,t+N_0K-1$. This analysis follows standard  arguments to analyze basic consensus algorithms (e.g., \cite{shiJSAC}), and we omit the details. The given claim therefore holds by computing the probability of the selection of the above sequence of arcs and the event $\{B_\tau=1,D_\tau=0,\tau=t,t+1,\dots,t+N_0K-1\}$. Note that  $N_0$ and $q_0$ depend on $\alpha,b_\ast,d_\ast,p_\ast,p^\ast,n$ but do not depend on $\beta$.

\vspace{2mm}

In addition, in view of Assumption A5, we can select a node $i_\ast \neq i_0$ satisfying  $(i_\ast,i_0) \in \bigcup_{\tau \in [t+N_0K,t+(N_0+1)K-1] } \mathcal{E}^-_\tau$. {  Consider the following sequence of events
\begin{align}
&\{s_{i_0}(t+N_0K)=M(t),s_{i_\ast}(t+N_0K) \geq {M(t)}/{2}\};\nonumber\\
&\{\exists \tau\in[t+N_0K,t+(N_0+1)K-1]\ \mbox{s.t.}\ (i_\ast,i_0)\in E^-_\tau\}; \nonumber\\
&\{D_\tau=1\}; \nonumber\\
&\{B_m=D_m=0,m\neq \tau\in[t+N_0K,t+(N_0+1)K-1]\}.
\end{align}
If they all happen then
\begin{align}
&|s_{i_0}(t+(N_0+1)K)|\nonumber\\
&= \Big|s_{i_0}(t+N_0K) -\beta\Big(s_{i_0}(t+N_0K)+s_{i_\ast}(t+N_0K) \Big)\nonumber\\
 &-\beta \sum_{j\in N^-_{i_0} (\tau)\setminus\{i_\ast\}}\Big( s_{i_0}(t+N_0K)+s_{j}(\tau)\Big) \nonumber\\
 &\alpha\sum_{j\in N^+_{i_0} (\tau)}\Big( s_{i_0}(t+N_0K)+s_{j}(\tau)\Big) \Big| \nonumber\\
&   \geq \Big(\frac{3}{2}\beta-1-\frac{n-2}{n}\Big) M(t),
\end{align}
where the last inequality is obtained from the facts that  $s_{i_0}(t+N_0K)=M(t)$, $\beta\Big(s_{i_0}(t+N_0K)+s_{i_\ast}(t+N_0K) \Big)\geq \frac{3}{2}\beta M(t)$,
$\beta \sum_{j\in N^-_{i_0} (\tau)\setminus\{i_\ast\}}\Big( s_{i_0}(t+N_0K)+s_{j}(\tau)\Big)\geq 0$, and
\begin{align}
\Big|\alpha\sum_{j\in N^+_{i_0} (\tau)}\Big( s_{i_0}(t+N_0K)+s_{j}(\tau)\Big) \Big|\leq \frac{(n-2)}{2n}\times 2M(t)
\end{align}
in light of $|N^+_{i_0} (\tau)|\leq n-2$ and $\alpha\leq (4n)^{-1}$.

It then follows that
\begin{align}
&\mathds{P}\Big(|s_{i_0}(t+(N_0+1)K)|\geq \big(\frac{3}{2}\beta-1-\frac{n-2}{n}\big) M(t) \Big)\nonumber\\
&\geq  \mathds{P}\Big( s_{i_0}(t+N_0K)=M(t),s_{i_\ast}(t+N_0K) \geq {M(t)}/{2}\Big)\nonumber\\
& \times  \mathds{P}\Big(  \exists \tau\in[t+N_0K,t+(N_0+1)K-1]\ \mbox{s.t.}\ (i_\ast,i_0)\in E^-_\tau \Big)\nonumber\\
& \times \mathds{P}(D_\tau=1) \nonumber\\
& \times \mathds{P}\Big(B_m=D_m=0,m\neq \tau\in[t+N_0K,t+(N_0+1)K-1]\Big)\nonumber\\
&\geq \vartheta_0, \nonumber
\end{align}
where $\vartheta_0=q_0 p_\ast d \big((1-d)(1-b)\big)^{K-1}$. This implies
\begin{align}\label{10}
\mathds{P}\big(M(t+(N_0+1)K)\geq  (3\beta-4) M(t)/2  \big) \geq \vartheta_0.
\end{align}

Now assume that $M(0)>0$ so that $U(m)=\log(M(m(N_0+1)K))$ for $m\ge 0$ is well defined. Note that from Lemma \ref{lem5} and (\ref{10}), we have:
$$
\mathds{E}\{U(m+1)-U(m)\} \ge -N_0K\log ({2n}) +  \vartheta_0 \log \big( ({3}\beta-4) /2\big).
$$
For $\beta$ large enough, the r.h.s. in the above inequality is strictly positive. We can then easily conclude, using classical arguments in random walks that the process $U(m)$ has a strictly positive drift, from which it can be deduced that $\mathds{P}\big( \liminf_{m\rightarrow \infty} M\big(mN_0K\big)=\infty \big)=1$ (for $\beta$ large enough). Using Lemma \ref{lem5}, one can easily conclude the desired theorem.

}




\subsection{Proof of Theorem \ref{prop1}}  The argument is based on the intuition that when one of the node states diverges, there is always a realization of edges with non-zero probability that ``pulls" another node toward divergence since these pulling actions happen infinitely often due to the Borel-Cantelli Lemma. Then suitable connectivity of the interaction graphs recursively leads to the desired no-survivor property.

Assume that  for some $q_\ast>0$ we have $\mathds{P}(\limsup_{t\rightarrow \infty} \max_{i\in \mathcal{V}} | s_i(t)| = \infty)\geq q_\ast$. There must be a node $i_0$ satisfying $\mathds{P}(\limsup_{t\rightarrow \infty}  | s_{i_0}(t)| = \infty)\geq q_\ast/n$. Let $C_0>0$, and define
$
T_1^\star:=\inf_{t}\big\{|s_{i_0}(t)| \geq C_0\big\}.
$
$T_1^\star$ is a stopping time. Let ${K}>0$ be an integer. We can further recursively define $T_2^\star,\dots,T_m^\star,\dots$ by
$$
T_{m+1}^\star:=\inf_{t\geq T_m^\star+{K}}\big\{|s_{i_0}(t)| \geq C_0\big\}.
$$
Based on  Theorem 4.1.3 in \cite{durr}, each $T_{m}^\star$ is a stopping time for all $m\geq 0$ and $({G}_{T_1^\star}, {B}_{T_1^\star}, {D}_{T_1^\star})$, $\dots$, $( {G}_{T_1^\star+{K}-1}, {B}_{T_1^\star+{K}-1}, {D}_{T_1^\star+{K}-1})$; $({G}_{T_2^\star}, {B}_{T_2^\star}, {D}_{T_2^\star}),\dots,( {G}_{T_2^\star+{K}-1}, {B}_{T_2^\star+{K}-1}, {D}_{T_2^\star+{K}-1});\dots$ are independent random variables that are also independent of $\mathcal{F}_{{T_1^\star}-1}$. In addition, we have $\mathds{P}( T_m^\star<\infty, m=1,2,\dots)\geq q_\ast/n$. Under Assumption A2, $\mathcal{G}\big({[T_1^\star,T_1^\star+K-1]}\big)$
being  strongly connected is a sure event. As a result, there exists another node $i_1\in\mathcal{V}\setminus {i_0}$ and $\tau_0\in [T_1^\star,T_1^\star+K-1]$ such that $(i_0,i_1)\in \mathcal{E}_{\tau_0}$. {Assume the event $s_{i_0} (\tau_0)=s_{i_0}(T_1^\star)$ (whose probability can be lower bounded by $B_t=D_t=0,t\in [T_1^\star,\tau_0-1]$ and is then  used to derive the $\chi_0$ below.) }We treat two cases: $\sigma_{i_0i_1}=-$ and $\sigma_{i_0i_1}=+$.
 \begin{itemize}
 \item[(i)] $\sigma_{i_0i_1}=-$.
 \begin{itemize}
\item If $\beta=1$, then $|\beta s_{i_0} (\tau_0) + (1-\beta) s_{i_1} (\tau_0)|= |\beta s_{i_0} (\tau_0)|=|s_{i_0}(T_1^\star)|  \geq {C_0}$;

\item  If $\beta\neq 1$ and $|s_{i_1} (\tau_0)|< {\beta C_0}/{(2|1-\beta|)}$, then $|\beta s_{i_0} (\tau_0) + (1-\beta) s_{i_1} (\tau_0)|\geq \beta C_0 -(1-\beta)|s_{i_1}(\tau_0)| \geq \beta C_0/2$.
\end{itemize}
\item[(ii)] $\sigma_{i_0i_1}=+$.
\begin{itemize}
\item If $\alpha=1$, then $|\alpha s_{i_0} (\tau_0) + (1-\alpha) s_{i_1} (\tau_0)|= C_0$.

\item  If $\alpha\neq 1$ and $|s_{i_1} (\tau_0)|<{\alpha C_0}/{(2|1-\alpha|)}$, then $|\alpha s_{i_0} (\tau_0) + (1-\alpha) s_{i_1} (\tau_0)|\geq \alpha C_0/2$.
    \end{itemize}
\end{itemize}
Now $s_{i_1}(\tau_0+1)=-\beta s_{i_0} (\tau_0) + (1-\beta) s_{i_1} (\tau_0)$ when $i_0$ is the unique node in  $N_{i_1}^-({\tau_0})$ and $D_{\tau_0}=1$. Also observe that
$s_{i_1}(\tau_0+1)=\alpha s_{i_0} (\tau_0) + (1-\alpha) s_{i_1} (\tau_0)$ when $i_0$ is the unique node in  $N_{i_1}^+({\tau_0})$ and $B_{\tau_0}=1$. Stationarity   ensures that $({B}_{T_1^\star}, {D}_{T_1^\star})$, $\dots$, $({B}_{T_1^\star+{K}-1}, {D}_{T_1^\star+{K}-1})$ have the same distribution as $({B}_{0}, {D}_{0})$. We can therefore simply bound the probabilities of the above events and establish
\begin{align}
\mathds{P}\big(\exists i_1\in \mathcal{V}\setminus\{i_0\} : |s_{i_1}(T_1^\star+K)|\geq \phi C_0\big)&\geq \chi_0, \nonumber
\end{align}
where $\chi_0=\big((1-b)(1-d)\big)^{2K-1} \min\{b,d\} p_\ast(1-p^\ast)^{n-2}$ and $\phi=\min \big\{ [ {\alpha }/{(2|1-\alpha|)}], \alpha/2, [{\beta }/{(2|1-\beta|)}], \beta/2,1\big\}$ (we use $[\cdot]$ to indicate that the corresponding term is taken into account in the $\min$ only if it is well defined). Repeating the analysis on $T_2^\star,\dots$ we obtain
\begin{align}
\mathds{P}\big(\exists i_m\in \mathcal{V}\setminus\{i_0\} : |s_{i_m}(T_m^\star+K)|\geq \phi C_0\big)\geq \chi_0. \nonumber
\end{align}
Since we have a finite number of nodes,  independence
allows us to invoke the second Borel-Cantelli Lemma (cf. Theorem 2.3.6 in \cite{durr}) and conclude that
\begin{align}\label{16}
&\mathds{P}\big(\exists \ i_1\in\mathcal{V}\setminus\{i_0\} : \nonumber\\
  &\ \ \limsup_{t\rightarrow \infty}|s_{i_1}(t)|\geq \phi C_0 \big|T_m^\star <\infty,m=1,\dots\big)=1.
\end{align}

Note that $C_0$ can be chosen arbitrarily, and hence (\ref{16}) implies that there exists $i_1\in\mathcal{V}\setminus\{i_0\}$ such that
\begin{align}\label{eq:cc}
\mathds{P}\big(\limsup_{t\rightarrow \infty}|s_{i_1}(t)|=\infty \big|\limsup_{t\rightarrow \infty} \max_{i\in \mathcal{V}} | s_i(t)| = \infty\big)=1.
\end{align}
We can apply the same argument recursively, to show that (\ref{eq:cc}) holds for any node $i_1$ in the network.

\section{Conclusions}
Inspired by examples from social, biological and engineering networks,  the emerging behaviors of node states  evolving over  signed random networks in a dynamical environment were studied. Each node received  positive and negative  recommendations from its neighbors determined by the sign of the  arcs. The positive recommendations were consistent with the standard consensus dynamics, while the negative recommendations flip the sign of node states in the local interactions as introduced by Altafini in \cite{altafini13}. After receiving recommendations, each node puts a deterministic  weight and a random attention on each of the recommendations and then updates  its state.  Various conditions were derived  regarding the almost sure convergence and  divergence of this model. These results have significantly  extended the analysis of the results of  \cite{altafini13} to more general models and detailed results.  The corresponding relative-state flipping   model \cite{shiJSAC, shiOR} under this general random graph model will be investigated in our future work.  Some other interesting future directions  include the co-evolution of the signs of the interaction links   along with the node states,  as well as the optimal placement of  negative links with the aim of breaking  the effect of positive updates as much as possible.

\section*{Acknowledgement}
This work has been supported in part by the Knut   and Alice Wallenberg Foundation, the Swedish Research Council,   KTH SRA TNG, and  by AFOSR MURI grant FA9550-10-1-0573.

\medskip

\medskip

\noindent {\sc Guodong Shi} \\
\noindent {\small College of Engineering and Computer Science, The Australian National University, \\ Canberra ACT 0200, Australia}\\  {\small Email: } {\tt\small guodong.shi@anu.edu.au}

\medskip

 \medskip

\medskip

\noindent {\sc Alexandre Proutiere, Mikael Johansson, and Karl H. Johansson} \\
\noindent {\small  ACCESS Linnaeus Centre,
   School of Electrical Engineering,
KTH Royal Institute of Technology,
\\ Stockholm 100 44, Sweden }\\
       {\small Email: 』 {\tt\small alepro@kth.se, mikaelj@kth.se, kallej@kth.se}

\medskip

\medskip

\noindent {\sc John S. Baras} \\
\noindent {\small Institute for Systems Research, University of Maryland,\\
College Park, MD 20742, USA}\\
{\small Email: } {\tt\small baras@umd.edu}

\end{document}